# A Survey on the Role of Artificial Intelligence in the Prediction and Diagnosis of Schizophrenia

Narges Ramesh[1], Yasmin Ghodsi[2], Hamidreza Bolhasani[1, *]

*Abstract*— **Machine learning is employed in healthcare to draw approximate conclusions regarding human diseases and mental health problems. Compared to older traditional methods, it can help to analyze data more efficiently and produce better and more dependable results. Millions of people are affected by schizophrenia, which is a chronic mental disorder that can significantly impact their lives. Many machine learning algorithms have been developed to predict and prevent this disease, and they can potentially be implemented in the diagnosis of individuals who have it. This survey aims to review papers that have focused on the use of deep learning to detect and predict schizophrenia using EEG signals, functional magnetic resonance imaging (fMRI), and diffusion magnetic resonance imaging (dMRI). With our chosen search strategy, we assessed ten publications from 2019 to 2022. All studies achieved successful predictions of more than 80%. This review provides summaries of the studies and compares their notable aspects. In the field of artificial intelligence (AI) and machine learning (ML) for schizophrenia, significant advances have been made due to the availability of ML tools, and we are optimistic that this field will continue to grow.**

## 1. Introduction

Schizophrenia is a complex mental disorder that affects how a person behaves, senses, speaks, and thinks, which can exceedingly reduce the quality of life if left untreated. Early diagnosis and functional treatment can improve the disease prognosis. Generally, it affects people ages 15 to 34. It's estimated that schizophrenia affects 24 million people worldwide, which is approximately 1 in 300 people (0.32%), and it's typically diagnosed in the early twenties for males and in the early thirties for females [1].

As of now, schizophrenia cannot be reliably diagnosed by a clinical test; instead, the patient's behavior and behavioral changes are the major factors in making the diagnosis. The patient's behavior is followed by significant alterations in mood, speech, thoughts, and behavior [2]. The symptoms fall into three major categories, which are commonly used positive and negative scales. Positive signs, are not normally experienced but are present in people during psychotic, such as hallucinations, experiencing things that are not real, delusions like paranoia, and disorganized thoughts and speech. Negative symptoms include shortages of regular emotional reactions, such as a loss of excitement during everyday activities and loss of motivation, and difficulty showing emotions and doing daily tasks. Cognitive deficits are the earliest and most commonly found symptoms in schizophrenia, like issues with concentration, memory, and attention. Since the cause of schizophrenia is not fully recognized, there isn't a sufficient treatment for it. Still, the symptoms can be managed and controlled using antipsychotic medications in combination with psychosocial interventions and social support.

As challenging as treating schizophrenia is, there are so many effective ways that can improve the symptoms and overall disease with the proper diagnosis and treatment. With the emergence of advanced technology and artificial intelligence, deep learning has greatly aided in the advancement of MRI research. In the realm of medical imaging, the implementation of deep learning has revolutionized the way we diagnose various medical conditions. Deep learning is a sub-type of machine learning that utilizes artificial neural networks (ANNs) to simulate the function of biological neurons by learning from vast amounts of data.

[1] Department of Computer Engineering, Islamic Azad University Science and Research Branch, Tehran, Iran.
[2] Department of Psychology and Educational Sciences, Islamic Azad University Central Tehran Branch
* Corresponding Author: hamidreza.bolhasani@srbiau.ac.ir

Artificial neural networks are designed to mimic the information processing and communication nodes of the brain system. They are composed of interconnected artificial neurons, which are organized layers. It is built around a network of interlinked components known as artificial neurons. Each neuron-to-neuron connection can transmit a signal, which the receiving neuron can then process and send to the other neurons it is linked to.

In the organization of this review, section 2 is focused on some fundamental concepts and a literature review about schizophrenia and deep learning. Section 3 contains a review of related studies and related works about the context of this survey. A summary table of related works is presented in this section. And at last, in section 4, a conclusion is presented.

## 2. Fundamental Concepts

In this section, fundamental concepts and a literature review related to this survey are covered. Firstly, symptoms and causes of schizophrenia are discussed and, in the following, methods of diagnosis and treatment are presented. Then the definition and mechanism of deep neural networks and their application in the prediction and diagnosis of schizophrenia using EEG signals and dMRI images are discussed. The studied papers are done between 2019 to 2022.

### 2-1. Schizophrenia Symptoms and Causes

In this section, there is an overview of clinical symptoms, causes, prevalence, diagnostic methods, and treatment of schizophrenia. The first discussion considered is the role of environmental and genetic factors in the early stages of life in changing the neurodevelopment pathways for a person's susceptibility to schizophrenia.

The first explanation that is of interest in this part is the role of environmental and genetic factors in the first stages of life to change the neurodevelopment pathways in a person's predisposition to schizophrenia. The; final topic considers the underlying mechanisms of schizophrenia treatment and the side effects of significant treatments [3].

Schizophrenia is associated with disabling a person in several areas and functional issues. There are many defining factors of disability that have been identified and found so far, including cognitive impairments, problems in daily practical and social skills, and difficulties in self-assessment abilities. Although psychotic symptoms have limited temporal and cross-sectional correlations with daily functioning, evidence suggests that long-term clinical stability, sometimes achieved through treatment with long-acting antipsychotics, can be improved by the daily performance of people is accompanied [4]. This review addresses the characteristics and factors of disability with therapeutic implications for each area and context of disability.

Episodic memory deficits are consistently a significant issue in the cognitive functioning of patients with schizophrenia, which are present from the beginning of the disease and are highly related to performance disability. In the past, researches using approaches such as episodic memory disorders were disproportionate, and cognitive neuroscience in schizophrenia has been investigated under relational encoding conditions [5]. In schizophrenia, a series of problems cause psychotic symptoms, including; abnormal function of midbrain dopamine to limbic areas. Considering this issue, recent advances in neuroimaging techniques have led to exciting and unexpected findings that investigate dopaminergic dysfunction in this disease [6].

## 2-2. Diagnosis and Treatment of Schizophrenia

Language deviation is one of the most common and prominent symptoms of schizophrenia. With the progress of linguistics, it is now possible to quickly check the language with exact criteria [7]. Schizophrenia symptoms can be different in different people, generally, there are three main categories: psychosis, negative/positive, and cognitive. The symptoms of psychosis and schizophrenia are as follows: Delusion, mental disorders, hallucination, and movement disorders.

There are a series of negative symptoms, loss of interest, loss of motivation to do enjoyable things or not enjoying day-to-day activities, isolation, difficulty in normal daily functioning, and difficulty in expressing emotions, and problems such as attention and focus are also included.

These symptoms can make it hard to learn new things or remember. A person's level of cognitive functioning is one of the best predictors of their daily functioning. People's cognitive functions are evaluated using special tests.

In schizophrenia and the psychological treatment of this disorder, people need long-term and complete treatment. Meanwhile, the role of the patient in the treatment process is significant, and individual differences in the treatment needs of people should be taken into account. The recent attention and interest in cognitive rehabilitation or training patients on how to manage cognitive deficits are promising. A series of family intervention programs to teach relatives and help them deal more effectively with the disease of these people have shown positive effects on the process of treating schizophrenia, but it appears that therapeutic gains are of uncertain duration [8]. There are a variety of treatments, such as psychological treatments, drug therapy, and biological treatments for schizophrenia, and even things like functional magnetic resonance imaging (fMRI) and Electroencephalogram (EEG) are done, which are explained here.

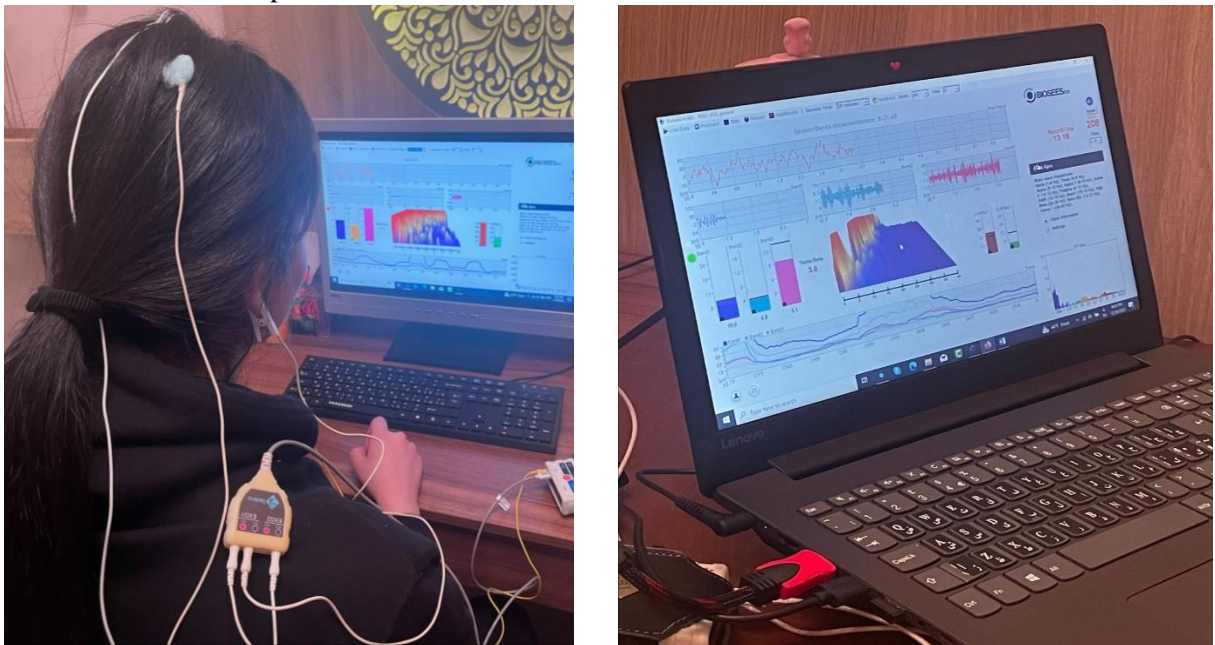

Fig.1. Left: EEG electrodes are attached to the patient's head so that we can see the brain waves clearly
Right: A view of the patient's brain waves during EEG

Magnetic resonance imaging has made a big difference in brain behavior and function. fMRI has been the first line of treatment for many reasons. In addition to saving time, fMRI has provided several other good functions with the help and application of neuroscience: higher spatial resolution, non-invasiveness, high reproducibility, and cost-effectiveness. As a result of this, a large number of fMRI studies have been conducted throughout the neurosciences, both in healthy individuals and in patients with brain disorders [9].

Timely and early diagnosis of schizophrenia can cause effective treatment and improve the patient's life as quickly as possible. In the past, many studies and research have been conducted using techniques such as EEG aimed at determining the neurobiological mechanisms and clinical symptoms that constitute schizophrenia. In addition to helping with more effective diagnoses, they can be very effective in predicting and anticipating the onset of schizophrenia [10]. An automatic, relatively inexpensive, and relatively accurate system for the diagnosing of schizophrenia is needed. Electroencephalography has been designed to evaluate and diagnose brain function. The role of the EEG method is to process brain states with different waves [11]. For a better understanding, pictures of a patient during EEG, the baseline, and the results are presented in Figures 1, 2, and 3 respectively.

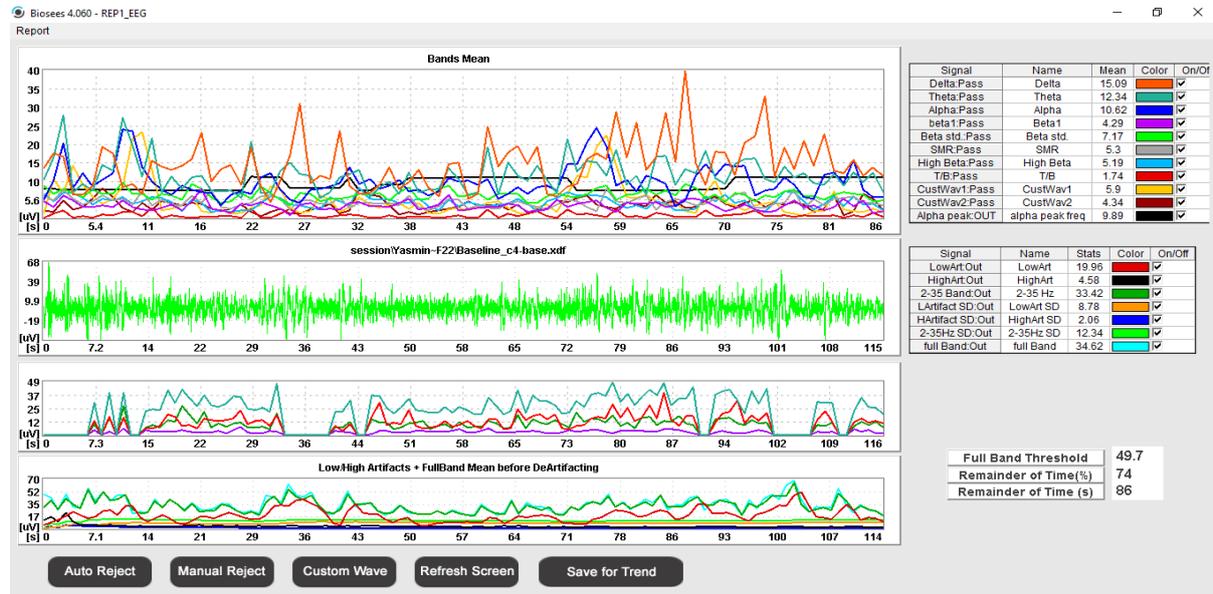

Fig. 2. EEG Baseline results with brain waves

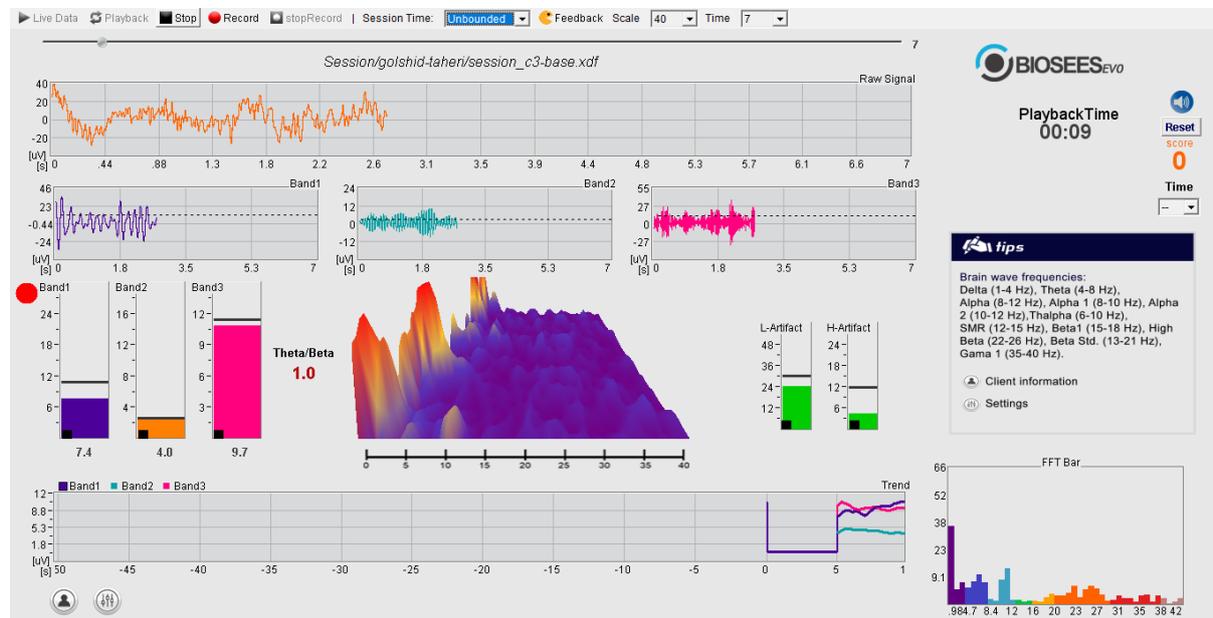

Fig. 3. During treatment and working with EEG

## 2-3. Deep Learning and Schizophrenia Diagnosis

Neural networks that possess multiple layers between their input and output layers are known as deep neural networks (DNNs). The term "deep" is used to indicate the presence of a large number of hidden layers. There can be as many as 150 layers in deep networks. Deep learning models are trained by using large sets of labeled data that learn features straight from the data, which makes it different from the traditional machine learning approach. Standard ML operates on extremely processed data. While traditional machine learning algorithms rely on pre-defined features, deep neural networks (DNNs) directly operate on high-dimensional raw data and extract optimal features through learning. This approach yields significant improvements in the performance and outcomes of deep learning as compared to conventional machine learning.

Convolutional neural networks (CNNs) are one of the most used deep neural networks [12], shown in Figure 4. CNN has got different layers, this includes a layer of convolution, a layer of non-linearity, a pooling layer, and a fully-connected layer. One of CNN's most vital features is its convolutional layer parameter sharing method, which lowers the number of parameters in ANN and makes it simpler and more effective to address complicated problems that were previously handled by class ANN. CNN does an outstanding job-solving machine learning problems, mostly in matters dealing with image data and its classification and computer vision.

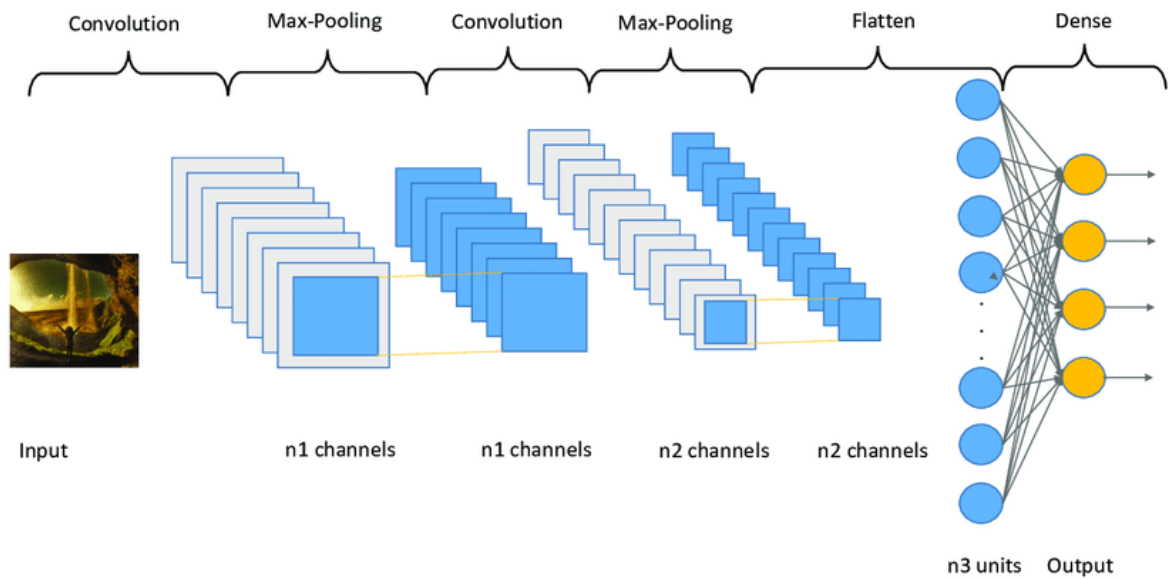

Fig. 4. Sample of a Convolutional Neural Network Architecture [13]

There are numerous ways in which deep learning can be used in detecting schizophrenia. Functional magnetic resonance imaging [13] investigates the difference between oxygenated and deoxygenated blood. fMRIs use the same technology as MRI. MRI produces images of organs and tissues while fMRI produces a photo showing the blood flow in the brain. This technology has improved neuroimaging in cognitive neuroscience, scanner technology, and image acquisition protocols.

fMRI analyses differences between different states of the brain [14]. Deep learning then uses its techniques and tools to detect baseline differences for identifying schizophrenia.

## 3. Review of Related Studies

Calhoun et al. [15] and Jafri and Calhoun [16] are among the finest works that imply fMRI data. According to their findings, when Schizophrenia patients and healthy controls were given the Sternberg working memory task and an oddball auditory test to complete, respectively. Images from 15 HC and 15 schizophrenia patients' fMRI scans revealed that schizophrenia patients seemed to "activate" less over a more limited number of distinct brain areas by switching up the test training sets. They were able to attain an average accuracy of 75.6% classification. Following this study, their subsequent work was improved by applying a multivariate analysis approach, which effectively distinguished patients with schizophrenia from those without schizophrenia with 95% specificity. This study inspired many other studies using fMRI data and deep learning techniques in classifying schizophrenia.

The study aimed to differentiate individuals with schizophrenia by utilizing recurrent neural networks on multi-site FMRI data. Weizhen Yan [17] and other researchers suggested a multi-scale RNN model using 558 patients with schizophrenia and 542 healthy controls.

Table. 1. An overview of the research and predictions using fMRI for schizophrenia diagnosis

| Reference | Year | Healthy Controls | Dataset | Successful Prediction | AI Technique |
|---|---|---|---|---|---|
| Weizheng Yan et al. [17] | 2019 | 542 HC | 588 patients | 83·2% | RNN |
| Jooyoung OH et al. [18] | 2019 | 72 HC | 72 patients | 98.09 ± 1.01% | 3D-CNN |
| Luca Steardo Jr et al. [19] | 2020 | 32 HC | 32 patients | Over 80% | SVM |

Traditionally EEG, which is a powerful tool, has been used for the diagnosis of mental disorders such as Schizophrenia and Parkinson's due to its easy setup and portable method. A few EEG-based machine learning techniques were used for feature extraction and implying classification methods for the detection of schizophrenia. In recent years, deep learning techniques had a significant impact on detecting brain signals and diagnosis of mental illnesses. EEG signals are controversial to be an effective tool for the diagnosis of schizophrenia due to their depth sensitivity, and they are not widely used in this field and highly rely on assumptions and prior knowledge about the patient's condition and symptoms. But this could be improved by the implementation of AI and ML [20].

Shu lih oh [21] and other scientists described a computerized identification of schizophrenia using a convolutional neural system. To analyze the signals, data was collected from 14 healthy individuals and 14 patients with schizophrenia. A convolutional neural network comprising 11 layers was used to analyze the data. The model demonstrated a subject-based testing accuracy of 98.07% and a non-subject-based testing accuracy of 81.26%.

In 2020, Siuly [22] and other authors worked on a computerized method for the automated detection of schizophrenia using EEG signals [23]. Their study aimed to provide a system for automatically identifying schizophrenia based on electroencephalogram (EEG) data. To diagnose schizophrenia using EEG data, they developed a method based on the empirical mode decomposition (EMD) methodology. The EMD technique was used in the study to split each EEG signal into Intrinsic Mode Functions (IMFs), and from these IMFs, twenty-two statistical characteristics/features were computed. Ninety controls and 90 schizophrenia patients participated in the research. Classifiers were employed to assess the generated feature sets for the EEG data set. The most accurate classifier has an accuracy rate of 89.59% for IMF and a rate of accurate classification for schizophrenia of 93.21%.

In 2021, Smith K. Khare [25] and other writers used SPWVD-CNN for the automatic identification of schizophrenia patients using EEG data to further the research. To make amends for the flaws of the feature extraction approach, the study authors suggested automated detection of schizophrenia utilizing a time-frequency analysis and convolutional neural network.

Table. 2. A summary of the research related to predictions of schizophrenia based on EEG

| Reference | Year | Healthy Controls | Dataset | Successful Prediction | AI Technique |
|---|---|---|---|---|---|
| Shu Lih Oh et al. [21] | 2019 | 14 HC | 14 Patients | 98.07% | CNN |
| Siuly et al. [22] | 2020 | 90 HC | 90 Patients | 93.21% | EMD |
| Jie Sun Et al. [24] | 2021 | 55 HC | 54 patients | 99.22% | EEG |
| Smith K. Khare et al. [25] | 2021 | 32 HC | 49 patients | 93.36% | Time-frequency analysis CNN |

Diffusion Magnetic Resonance Imaging (dMRI) is a method used by neuroscientists for detecting information about structural connections inside the brain. This technique uses the creation of image contrast through the movement of water molecules in tissue. The diffusion tensor describes a wide range of quantities, which includes magnitude, anisotropy degree, and direction of the diffusion anisotropy. Perfusion MRI (pMRI) is a technique widely used in measuring cerebral perfusion with the help of a variety of homodynamic metrics, such as blood volume and cerebral blood flow, and means transit time which is used mainly in treating individuals who suffer from cerebrovascular disease or other brain disorders, such as schizophrenia. Due to its ability to monitor blood flow, PMRI is highly helpful in determining how well pharmacological treatments for schizophrenia work. In 2018 Harvard Review of Psychiatry worked on The Amygdala in Schizophrenia and Bipolar Disorder. Widely used in psychotic spectrum disorders, the amygdala contributes to the understanding of Schizophrenia's convergent and divergent neurological bases. A systematic search has been done since 2017 to discover the amygdala neuroimaging studies for Schizophrenia and a few other mental disorders, such as bipolar disorder, focused on diffusion tensor imaging (DTI). In this review, they chose ninety-four studies. The DTI studies indicated a decrease in fractional anisotropy (a sign of abnormal white matter microstructure).

White matter damage in patients with schizophrenia; was the subject of Sung WooJoo's and his team's 2021 work on the dMRI data, which was yet another spectacular achievement: A multi-site diffusion MRI study [27]. There have been several multi-site diffusion magnetic resonance (dMRI) investigations done for a better statistical outcome because the evidence for white matter abnormality is insufficient in Schizophrenia. Using raw dMRI data from 190 Schizophrenia patients and 242 healthy controls, the scientists created a method.

They compared the diffusion measurements between the two groups, examined the impact of age and sex, and evaluated the relationship between fractional anisotropy (FA) and the symptoms. Significant differences were observed in fractional anisotropy (FA) in several tracts, including the right superior longitudinal fasciculus (SLF) iii, the left lateral orbitofrontal commissural tract, and the pOr-pOr commissural tract, as well as the pTr-pTr commissural tract. The right SLF iii FA and the positive and negative PANSS ratings were linked in the patient group with the pTr-pTr commissural tract FA. In their most recent studies, they found that patients with Schizophrenia had a decreased FA of the right SLF iii. In earlier investigations, it was discovered that Schizophrenia had a lower FA of the SLF.

In 2022, Matthew J. Hoptman [26] and other researchers looked at the connections between Resting-State Functional Connectivity and Diffusion Tensor Imaging in Patients with Schizophrenia and

Healthy Controls. Using resting state fMRI and diffusion tensor imaging, they examined 25 healthy control volunteers and 29 patients with Schizophrenia. Functional and Tractography The default mode network's functional and structural connections were estimated using the Connectivity Analysis Toolbox (FATCAT). Out of the 28 available area pairings, the relationship between modalities was examined, and 9 of them revealed tracts that were consistent (>80%) across participants. They discovered that FATCAT could be an excellent technique for analyzing the connection between imaging modalities.

Table. 3. A summary of the studies related to the role of dMRI in schizophrenia diagnosis using AI

| Reference | Year | Healthy Controls | Dataset | Successful Prediction | AI Technique |
|---|---|---|---|---|---|
| Sung WooJoo et al [27]. | 2021 | 242 HC | 190 Patients | - | Slice Diffusion QC |
| Matthew J. Hoptman et al [26]. | 2022 | 25 HC | 29 Patients | Over 80% | Functional and Tractography Connectivity Analysis Toolbox (FATCAT) |

## 6. Conclusion

The objective of this study is to detect and classify schizophrenia by leveraging AI techniques and machine learning algorithms to detect schizophrenia. These techniques leverage complex algorithms and computational models to analyze and quantify patterns in large sets of data, such as brain imaging scans, that can be indicative of the presence of the illness. In this paper we conducted a literature review of artificial intelligence and deep learning, highlighting their improvements and advances in the diagnosis of schizophrenia. This paper gives an overview and comparative analysis of studies conducted in this area within the last five years, which have utilized EEG signals, functional magnetic resonance imaging (fMRI), and diffusion magnetic resonance imaging (dMRI) to investigate the detection and classification of schizophrenia. These studies collected data from both healthy controls and patients with schizophrenia. All research conducted in this area has demonstrated classification accuracy results of over 80% for detecting schizophrenia. It is expected that this field will continue to grow and achieve more advancements in the future.

**Abbreviations**

| ANN | Artificial Neural Network |
|---|---|
| MRI | Magnetic Resonance Imaging |
| fMRI | Functional Magnetic Resonance Imaging |
| dMRI | Diffusion Magnetic Resonance Imaging |
| pMRI | Perfusion MRI |
| ML | Machine Learning |
| DL | Deep Learning |
| EEG | Electroencephalogram |
| EMD | Empirical Mode Decomposition |

| IMF | Intrinsic Mode Function |
| CNN | Convolutional Neural Network |
| DTI | Diffusion Tensor Imaging |

**Authors Biography**

**Narges Ramesh**

Narges Ramesh has a bachelor's degree in Computer Engineering from the University of Science and Research in Tehran. After completing her degree, she started working as a data analyst at Iran's largest VOD company. She has always been fascinated by deep learning and its applications in the medical field, particularly in the diagnosis and treatment of chronic illnesses. She wishes to continue studying and conducting research in this area.

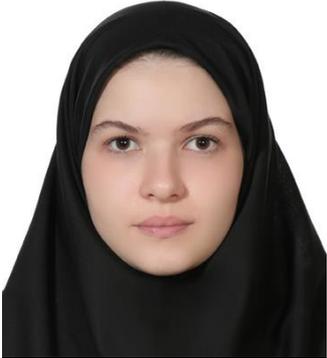

**Yasmin Ghodsi**

Yasmin Ghodsi received her BSc in counselling from the Islamic Azad University of Central Tehran Branch in 2021 and is currently studying MSc Clinical Psychology at the same university. She is working in the Yasmina Psychology and Psychiatry Clinic. She is very interested in the fields of Neuroscience, Artificial intelligence, and Psychology, and she has completed several courses in these fields. She is enthusiastic about science and learning.

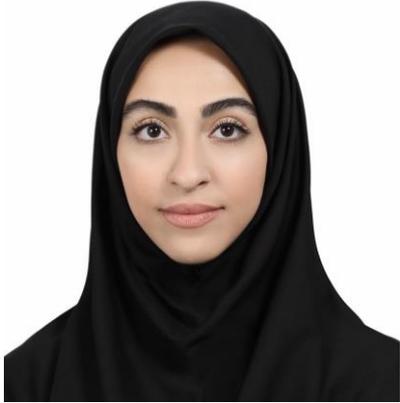

**Hamidreza Bolhasani, PhD**

AI/ML Researcher / Visiting Professor

Founder and Chief Data Scientist at DataBiox

Ph.D. Computer Engineering from Science and Research Branch, Islamic Azad University, Tehran, Iran. 2018-2023.

Fields of Interest: Machine Learning, Deep Learning, Neural Networks, Computer Architecture, Bioinformatics

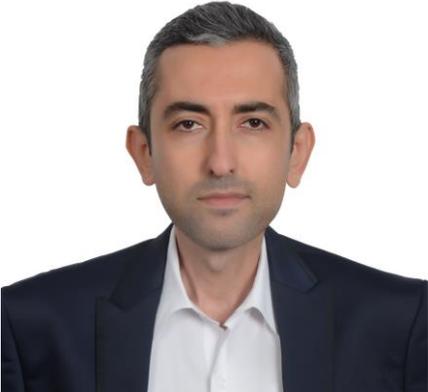